\documentclass[pdflatex,sn-mathphys-num]{sn-jnl}% Math and Physical Sciences Numbered Reference Style
\usepackage{graphicx}
\usepackage{multirow}
\usepackage{amsmath,amssymb,amsfonts}
\usepackage{amsthm}
\usepackage{mathrsfs}
\usepackage[title]{appendix}
\usepackage{xcolor}
\usepackage{textcomp}
\usepackage{manyfoot}
\usepackage{booktabs}
\usepackage{algorithm}
\usepackage{algorithmicx}
\usepackage{algpseudocode}
\usepackage{listings}
\usepackage{xcolor}
\usepackage{lmodern}
\usepackage{verbatim}

\begin{document}

\title[Article Title]{Magnetic Brightening and Nanoscale Imaging of Spin-Polarized Helical Edge Modes}


\author[1,2]{\fnm{Samuel} \sur{Haeuser}}

\author[1]{\fnm{Richard H. J.} \sur{Kim}}
\author[1,2]{\fnm{Lin-Lin} \sur{Wang}}

\author[1]{\fnm{Thomas} \sur{Koschny}}

\author[3,4]{\fnm{Pedro M.} \sur{Lozano}}

\author[3]{\fnm{Genda} \sur{Gu}}

\author[1,2]{\fnm{Randall K.} \sur{Chan}}

\author[1]{\fnm{Joong-Mok} \sur{Park}}

\author[1]{\fnm{Martin} \sur{Mootz}}

\author[1]{\fnm{Liang} \sur{Luo}}

\author[3,4]{\fnm{Qiang} \sur{Li}}


\author*[1,2]{\fnm{Jigang} \sur{Wang}}\email{jgwang@ameslab.gov; jgwang@iastate.edu}

%\affil*[1]{\orgdiv{Department}, \orgname{Organization}, \orgaddress{\street{Street}, \city{City}, \postcode{100190}, \state{State}, \country{Country}}}

%\affil[2]{\orgdiv{Department}, \orgname{Organization}, \orgaddress{\street{Street}, \city{City}, \postcode{10587}, \state{State}, \country{Country}}}

%\affil[3]{\orgdiv{Department}, \orgname{Organization}, \orgaddress{\street{Street}, \city{City}, \postcode{610101}, \state{State}, \country{Country}}}

\affil[1]{\orgdiv{Ames National Laboratory}, \orgname{U.S. Department of Energy}, \orgaddress{\city{Ames}, \postcode{50011}, \state{IA}, \country{USA}}}

\affil[2]{\orgdiv{Department of Physics and Astronomy}, \orgname{Iowa State University}, \orgaddress{\city{Ames}, \postcode{50011}, \state{IA}, \country{USA}}}

\affil[3]{\orgdiv{Condensed Matter Physics and Materials Sciences Department}, \orgname{Brookhaven National Laboratory}, \orgaddress{ \city{Upton}, \postcode{11973}, \state{NY}, \country{USA}}}

\affil[4]{\orgdiv{Department of Physics and Astronomy}, \orgname{Stony Brook University}, \orgaddress{ \city{Stony Brook}, \postcode{11974}, \state{NY}, \country{USA}}}


\abstract{Efficient sub--10~nm electric transport remains a major challenge for nanoelectronics due to high losses and impedance mismatches in conventional Drude metals. Despite their promise of dissipationless, reflection-free conduction, topologically protected chiral edge modes remain little explored in their nanoscale spin-polarized transport--particularly regarding real-space visualization, magnetic-field tunability, and high-frequency edge conductivity.
Here, we report magnetic brightening and nanoscale visualization of highly spin-polarizable infrared helical edge states using cryogenic magneto-infrared scattering-type scanning near-field optical microscopy (cm-IR-sSNOM).
Our measurements reveal magnetic-field–induced near-field conductivity at step edges, uncovering quantum spin Hall spin-splitting modes with enhanced infrared polarizability and slightly narrowed near-field profiles.
In addition, the infrared edge electrodynamic response scales nearly linearly with atomic layer number, providing compelling evidence that magnetic-field-induced gaps do not disrupt individual-layer edge states at energies of $\sim$100~meV.
These results sharply contrast with microwave and DC transport, where even small magnetically induced gaps decrease edge conduction. 
Magnetically tunable, topologically robust high-frequency edge modes open a pathway toward ultralow-loss nanoscale interconnects and quantum logic architectures for next-generation microelectronics, spintronics and quantum information science.}

\maketitle

\section{Introduction}\label{sec1}

A defining feature of quantum spin Hall (QSH) insulators is the presence of two counterpropagating edge modes with opposite spin orientations, giving rise to spin-polarized conduction currents \cite{hasan, Qi}. These states uniquely offer dissipationless transport even at single-nanometer scales, defying traditional limitations. QSH insulator's intrinsic spin-momentum locking that is protected by time-reversal symmetry, regardless of inversion symmetry, prevents backscattering, effectively eliminating impedance mismatches in sophisticated electronic circuits. However, despite their advantages, experimental observations at DC and microwave frequencies reveal that external magnetic fields introduce significant suppression that is attributed to magnetic field induced gap opening and inter-layer hybridization~\cite{shiMIM, Jiang2023}. 
The magnetic suppression challenge observed in lower-energy probes underscores the need for investigations of chiral edge transport at higher frequencies, well beyond the hybridization gap. However, direct infrared real-space visualization of QSH edge conductivity, as illustrated in Figs.~1a-1b, from spin-split bands---particularly under magnetic fields and cryogenic temperatures at Tesla and Kelvin scales---remains elusive, leaving a critical gap in our ability to understand and control these high frequency edge modes.

Zirconium pentatelluride (ZrTe$_5$) occupies a unique position at the boundary between strong and weak topological insulating phases. 
Distinct spin-polarized density of states \cite{Chen2015}, localized along crystal step edges (SEs) or domain walls (DWs) and controllable by magnetic fields, as illustrated in Figs.~1c and 1d, have been probed using scanning tunneling microscopy (STM).
For example, prior STM studies of monolayer and few stacking layers in ZrTe$_5$ have revealed intricate QSH transport characteristics, confirming the proximity of the bulk band gap to the SEs ~\cite{LiPRL, wuPRX}. These studies show the ubiquitous presence of topological edge states that are protected by spin-momentum locking and time-reversal symmetry~\cite{WengPRX, kons, kim, luo, vasw2, shiMIM}. Notably, the high-energy states in ZrTe$_5$ retain their topological robustness even under relatively strong magnetic fields, owing to strong spin-orbit coupling that persists well above the magnetic hybridization gaps at terahertz frequencies and continues to govern their electronic character. While STM has revealed spin-dependent band splitting and local density-of-states in ZrTe$_5$ under applied magnetic fields, it lacks the contrast needed to resolve the electrodynamics and counterpropagating AC currents--key signatures encoded in the phase-sensitive, complex infrared near-field responses of these spin-split states.

Cryogenic-magneto infrared/THz scattering-type scanning near-field optical microscopy (cm-THz-sSNOM) probes the phase-sensitive, oscillatory electric fields governed by the complex conductivity in nanometer-scale helical edge channels under high magnetic fields and sub-2K cryogenic temperatures \cite{kim3}. Figure~1a shows a representation of distinct conducting spin channels of ZrTe$_5$ under magnetic field. By polarizing high-frequency edge modes beneath the AFM probe tip, sSNOM signals directly resolve infrared photocurrent oscillations through combined amplitude and phase contrast. As shown in Figs.~1c–1d, This capability enables direct visualization of a unique magnetic brightening arising from imbalanced, spin-dependent edge channels at the nanometer scale—distinct from the magnetic quenching typically observed in microwave and DC transport studies.
Specifically, in zero magnetic field, spatially overlapping edge modes with opposite momenta ($\mathbf{k}$ and $-\mathbf{k}$) and spins (Fig.~1e) generate two AC edge mode flows with a $\pi$ phase difference, which largely cancel under the driving laser field and suppress the IR near-field contrast. Applying a magnetic field breaks the symmetry of the spin-dependent electronic dispersion near $E_\text{F}$ (Fig.~1a) and spatially separates the spin-polarized edge modes. As illustrated at intermediate (Fig.~1f) and high (Fig.~1g) field strengths, this field-induced asymmetry lifts the degeneracy of counter-propagating edge modes, creating imbalanced AC conducting currents resulting in a net current $J$ under laser driving, as illustrated for SEs  (Fig.~1c) and DWs  (Fig.~1d). Thus, the magnetic-field–induced changes give rise to complex infrared conductivity contrasts, providing sensitivity to spin-splitting energies and spatial narrowing of the edge modes, as directly probed by cm-IR-sSNOM signals at $\sim 100$~meV. Interestingly, the DWs (Fig.~1d) promote the formation of counter-propagating edge currents on both sides, originating from looped electron trajectories that encircle the outer edges of adjacent ZrTe$_5$ strips. These distinguishing features thereby enable direct visualization of markedly pronounced edge near-field signals induced by magnetic field and counterpropagating mode flow via asymmetric and phase-sensitive electric-field scattering. However, neither the magnetic brightening of spin-polarized edge-mode transport nor their real-space visualization has previously been observed in sSNOM signals, largely due to the technical challenges of reaching sufficiently low temperatures and high magnetic fields.

\begin{figure}[!ht] 
\centering 
\includegraphics[width=\textwidth]{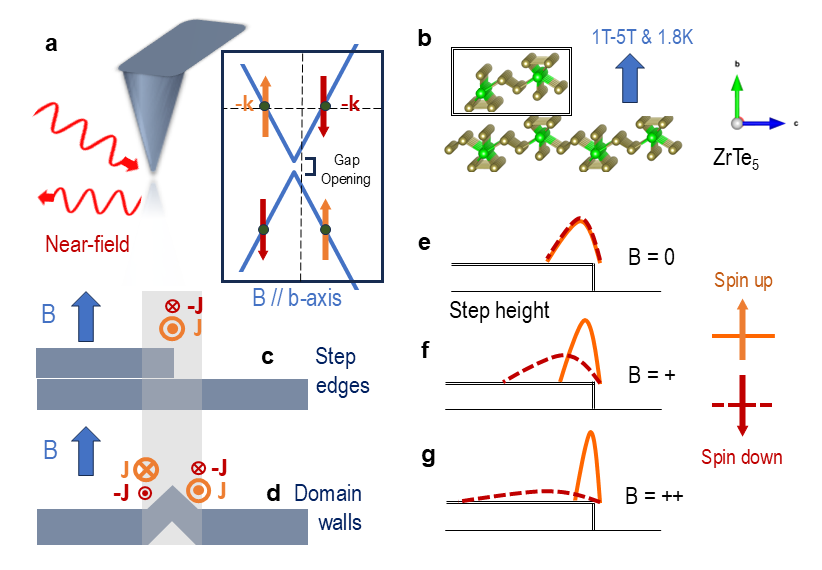} 
\caption{\textbf{Cryogenic magnetic infrared nano-imaging of spin-polarized edge states at crystal strip domains in ZrTe\textsubscript{5}.} 
\textbf{a}, Representative schematic of an atomic force microscope (AFM) probe scattering infrared near-field signals above a ZrTe\textsubscript{5} sample, operated at 1.8~K and up to 5~T. When a finite magnetic field is applied, the spin degeneracy of the edge Dirac bands is lifted, resulting in a small magnetic gap and a spin imbalance at the Fermi energy, $E_\mathrm{F}$, as illustrated in the right panel.
\textbf{b}, A magnetic field (blue arrow) is applied perpendicular to the ZrTe\textsubscript{5} sample surface along the crystallographic \textit{b}-axis, shown here for a single step edge.  
\textbf{c}, Schematics of step edges formed by stacking a few single layers. 
\textbf{d}, Schematics of domain walls arising from stacking variations or bending of layers meeting at a sharp boundary. 
Under laser driving, QSHE spin-splitting characteristics manifest as the conducting channels carrying net current $J$ that strongly depend on edge geometry. Step edges (\textbf{c}) exhibit a uniform net current direction, whereas domain walls (\textbf{d}) support opposite net currents on either side of the boundary.
\textbf{e-g}, Illustration of the real-space distribution of spin-polarized, counter-propagating edge modes at magnetic fields \( B = 0 \) (e), an intermediate field (f), and a high field (g)--highlighting the evolution of edge-state configurations with increasing magnetic field strength. As the field increases, the spin-up helical edge state (orange solid line) becomes increasingly confined to the step-edge boundary, while the spin-down counterpart (red dashed line) progressively spreads toward the inner bulk region \cite{LiPRL}.
 } 
\label{fig1:main}
\end{figure}

In this Article, we demonstrate, for the first time, magnetic near-field imaging at sub–liquid-helium temperatures (down to 1.8 K) to probe nanoscale spin-polarized edge modes, directly visualizing and controlling magnetic-field–brightened helical edge currents in the topological insulator ZrTe$_5$. 
%The robust, magnetically polarized edge modes are uniquely captured through comprehensive cryogenic magneto-infrared nano-imaging, spanning temperatures down to 1.8~K and magnetic fields up to 5~T. %Furthermore, our measurements reveal that edge conduction channels scale nearly linearly with the number of atomic layers, implying minimal hybridization across layers at infrared energies (~100 meV). 
%Unlike microwave and DC transport, where small magnetic gaps readily suppress edge conduction,
The infrared near-field edge responses intensify with increasing magnetic field and scale nearly linearly with atomic layer number, revealing an energy-scale–dependent topological robustness. Specifically, individual layers preserve their discrete edge transport identities at $\sim$100~meV energies, sharply contrasting with DC and microwave studies, where even small magnetic gaps quench edge conduction. 

\section{Results}\label{sec2}

To perform real-space measurements and differentiate counter-propagating edge modes, our cm-IR-sSNOM setup is based on a tapping-mode atomic force microscope (AFM) inside a top-loading 5-Tesla split-pair magnet cryostat with a base temperature of 1.8~K through a helium exchange gas system (Methods). We focus mid-infrared electromagnetic fields ($\sim$106-116~meV respectively) to an AFM tip that acts as an antenna to both focus the incident light and amplify the near-field scattering (Fig.~1a) \cite{htchen, Eisele2014, hillenbrand2025visible, Kim_new, kim3, kim, kim2, kim4, kim5, fei, hill, ribb, mast, von, dapo, dapolito, jutopo,guo2024terahertz}. Quasi-heterodyne detected near-field signals are extracted from the scattered electromagnetic waves demodulating from the tip--sample system at the $n_\text{th}$ harmonics of the tip-tapping frequency of the AFM. By interfering the complex scattered field $E_{\mathrm{NF}}^{(n)}$ with a reference beam, the phase-sensitive signal, $S_{n\Omega}$ at the $n_\text{th}$ harmonic of the tip oscillation frequency is obtained.
%proportional to the product of the reference field $E_{\mathrm{ref}}$, the complex near-field harmonic amplitude $E_{\mathrm{NF}}^{(n)}$, and a phase factor $e^{i\phi^{(n\Omega)}_\mathrm{NF}}$.
% such that $S_{n\Omega} \propto E_{\mathrm{ref}} \left| E_{\mathrm{NF}}^{(n)} \right| e^{i\phi_{\mathrm{NF}}^{(n\Omega)}}$.  
%$S_{n\Omega} = E_{\mathrm{ref}} \left| E_{\mathrm{NF}}^{(n)} \right| e^{i\left( \phi_{\mathrm{ref}} - \phi_{\mathrm{NF}}^{(n)} \right)}$. 
Unwanted far-field effects are suppressed while preserving a reasonable signal-to-noise ratio by projecting all phase-sensitive signals at $n=3$ or $4$ modulation, thereby enabling direct measurement of direction-dependent, polarized AC current flow from the tip-scattered complex near field.

\begin{figure}[!ht] 
\centering 
\includegraphics[width=\textwidth]{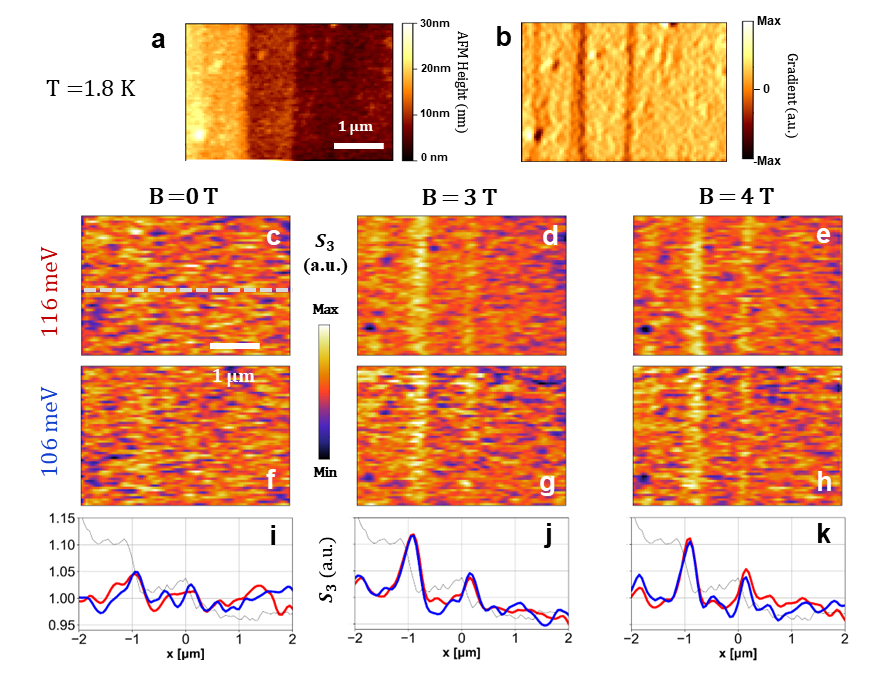} 
\caption{\textbf{Magnetic brightening revealed by near-field signals acquired at 1.8~K under varying magnetic fields in a representative layered region of ZrTe$_5$}
\textbf{a}, AFM topography featuring two step edges and \textbf{b}, horizontal Sobel gradient of the AFM image showing sharp sloped boundaries. 
\textbf{c-e}, cm-IR-sSNOM images taken at 10.652~$\mu$m ($\sim$116~meV) for 0~T, 3~T, and 4~T respectively. 
\textbf{f-h}, cm-IR-sSNOM images taken at 11.262~$\mu$m ($\sim$106~meV) for 0~T, 3~T and 4~T respectively.
\textbf{i-k}, Line cuts of AFM topography (black) and SNOM images at 10.652~$\mu$m (red) and 11.262~$\mu$m (blue) across the two step edges with heights of 5~nm and 9~nm for 0~T, 3~T and 4~T, respectively, at the gray dashed line shown in \textbf{c}.} 
\label{fig2:main}
\end{figure}

Fig.~2 presents typical near-field images acquired at various magnetic fields, with the cm-IR-sSNOM setup cooled to a base temperature of 1.8~K. We focus on a clean ``staircase" like structure consisting of the typical stacking step edges of ZrTe$_5$ layers. The AFM topography and the Sobel gradient in the scanning direction of the region of interest are shown in Fig.~2a-2b. 
The AFM images reveal step heights of approximately 9~nm and 5~nm from left to right, corresponding to stacking of 11 and 6 atomic layers, respectively. The Sobel gradient highlights the two distinct step-edge boundaries. 
Interestingly, the infrared near-field images reveal a pronounced magnetic-field-induced edge-brightening effect, as illustrated in Fig.~1c, consistently observed at two probe wavelengths. Figures~2c–2e show cm-IR-sSNOM images at 0~T, 3~T, and 4~T, respectively, acquired at an excitation wavelength of 10.652~$\mu$m. Figures~2f–2h present the corresponding cm-IR-sSNOM images at a different excitation wavelength of 11.262~$\mu$m, exhibiting the same magnetic-field dependence. An enhancement of the near-field signal is correlated with the stacking-layer boundaries shown in Figs.~2a–2b. In particular, increasing magnetic fields markedly amplify the edge signals, forming well-defined conducting edge channels visible in Figs.~2c–2h. From left to right, two distinct edge enhancements are observed, with the higher step edge (left) showing a stronger response than the lower step edge (right). All images are demodulated to the third harmonic, $S_3$.
%We follow the same process as described for the 60 K images and plot line cuts for both 10.652 $\mu$m (red) and 11.262 $\mu$m (blue) against the AFM topography (black). 
The magnetic brightening of the edge-conducting channels is further detailed in Figs.~2i–2k as line cuts of the cm-IR-sSNOM images, taken along the gray dashed line in Fig.~2c, at 10.652~$\mu$m (red) and 11.262~$\mu$m (blue) for 0~T, 3~T, and 4~T, respectively. At 0~T [Fig.~2i], the line cut shows only slight contrast at positions corresponding to the two step-edge features in the AFM topography line cut (gray). With increasing magnetic field [Figs.~2j–2k], well-defined peaks significantly enhance, indicating the formation of polarized edge conductive modes with strong near-field contrast along the step edges. This conclusion is fully consistent with the spin-population imbalanced edge modes that lead to a net near-field electromagnetic response, as illustrated for SEs (Fig.~1c).

\begin{figure}[!ht] 
\centering 
\includegraphics[width=\textwidth]{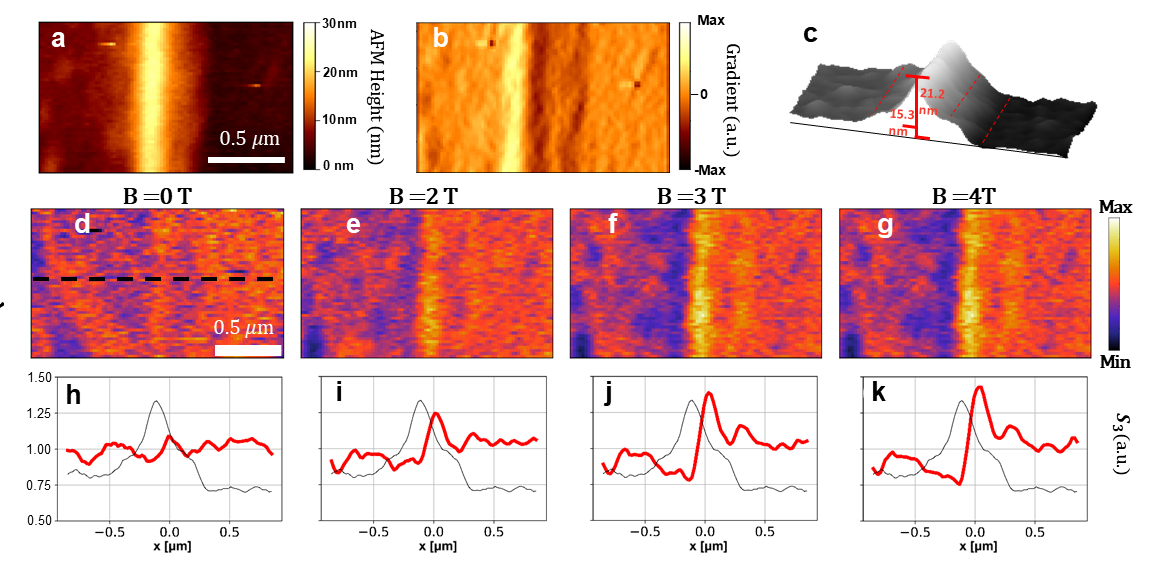} 
\caption{\textbf{Near-field images acquired across a sharp domain boundary in ZrTe$_5$ at 60~K under varying magnetic fields.}
\textbf{a}, AFM topography of the sample surface.  
\textbf{b}, Horizontal Sobel gradient of the AFM image highlighting two flat plateaus separated by sloped regions that converge at a sharp domain boundary.
\textbf{c}, 3D representation of AFM topography. Red dashed lines indicate regions of interest and mark the height variations of key surface features.
\textbf{d--g}, SNOM images acquired at 10.652~$\mu\text{m}$ ($\sim 116$~meV) under magnetic fields of 0~T, 2~T, 3~T, and 4~T, respectively.
\textbf{h--k}, Line profiles of AFM topography (black) and SNOM signal (red) corresponding to 0~T, 2~T, 3~T, and 4~T, respectively, taken along the black dashed line indicated in \textbf{d}.
} 
\label{fig3:main}
\end{figure}

A distinctive manifestation of magnetic-field-induced spin-split edge modes is the emergence of two spatially separated, counterflowing electron channels on opposite sides of the domain wall, which become increasingly pronounced with rising magnetic field.
%One important regime of study is the Lifshitz transition point of ZrTe$_5$. Previous studies of our sample show this transition temperature to occur at approximately 60 K. 
To observe this behavior, we identify two distinct layers of ZrTe$_5$ separated by a domain boundary, as illustrated in Fig.~1d. At this location, the AFM topography [Fig.~3a] and the Sobel gradient [Fig.~3b] clearly reveal its existence and structure. A 3D topographic view of this region [Fig.~3c] highlights three distinct locations, marked by red dashed lines across the ``hill." Near-field images acquired at various magnetic fields at 60~K show strong contrast at these sharp topographic edges with distinct magnetic field dependence. Figures~3d–3g present cm-IR-sSNOM images at 0~T, 2~T, 3~T, and 4~T, respectively, which exhibit magnetic brightening with opposite polarity. Similar to the step edges in Figs.~2c, 2f and 2i, the 0~T sSNOM image in Fig.~3d shows only slight contrast at positions corresponding to topographic features. As the magnetic field increases in Figs.~3e-3g, the near-field signals are strongly amplified, with opposite polarity on the left (dark) and right (bright) sides of the hill. This asymmetric near-field scattering is fully consistent with counter-propagating IR edge currents on opposite sides of the DWs as illustrated in Fig.~1d. They arise from field-induced spin population imbalances in the infrared polarizability responses. The resulting AC near-field polarizes the scanning probe tip dipole, which subsequently emits to the far field and is detected via phase-sensitive, near-field signals scattered from an asymmetric tip dipole. 
% $\left| E_{\mathrm{NF}}^{(3)} \right| e^{i\phi_{\mathrm{NF}}^{(3\Omega)}}$ shown in the plots.
%In these measurements, the AFM probe is excited at a wavelength of 10.652~$\mu$m (116~meV), the tip-tapping frequency is demodulated to the third harmonic $S_3$, and the forward and backward scans are averaged, denoted $S_3^{\mathrm{fwd+bwd}}$.

To highlight the magnetic-field-induced edge signals, Figs.~3h–3k present line cuts across the domain boundary, taken along the black dashed line in Fig.~3d. The near-field signal line cuts (red) exhibit an “inductive” line shape, with both negative and positive peaks, overlaid on the AFM topography line cuts (black). Each trace is normalized by dividing by its overall mean, so the plotted values represent deviations from this mean. The primary peak (strongest response) increases by 6~\%, 25~\%, 35~\%, and 48~\% from the mean for 0~T, 2~T, 3~T, and 4~T, respectively. A secondary, less-bright peak follows the jagged boundary between the intermediate step and lower plateau, increasing by 3~\%, 10~\%, 15~\%, and 20~\% for the same fields. On the opposite (left) side of the boundary, the contrast is inverted, with decreases of $-5$~\%, $-10$~\%, $-22$~\%, and $-28$~\% for 0~T, 2~T, 3~T, and 4~T, respectively. While the exact values vary with the position of the line cut, the images consistently show an increasing trend on the right side of the “hill” and a decreasing trend on the left side with increasing magnetic field---a pattern robust across the entire sSNOM images.

To quantitatively account the spin splitting of edge modes induced by the B field, Figure~4a summarizes the average positive (negative) deviations along the edge boundaries for the cm-IR-sSNOM images in Figs.~3h–3k. We compare these deviations for both polarities (red triangles and inverted blue triangles) with the average QSH edge-state spin-dependent energy splitting measured in prior STM work~\cite{LiPRL}. The sSNOM results show a good agreement with the STM data: at lower magnetic fields, the edge-state effect is minimal, whereas above $\sim 2$~T, the energy splitting associated with the 1D QSH effect emerges in both datasets. This correspondence provides compelling evidence for tip-induced spin-polarized infrared photocurrent transport localized at step edges, leading to the observed magnetic-field-induced brightening.

\section{Discussions}\label{sec3}
%We attribute the opposing magnetic field induced effects to the electronic field polarizing the tip dipole with $\chi_{cross}$, resulting in tip polarization emission to the far field. 
%Additionally, it's important to note that the near field will also have a symmetric contribution due to the oscillatory light field polarizing the tip and cantilever, which fully symmetrically polarizes the edge channels on domain walls.

\begin{figure}[!ht] 
\centering 
\includegraphics[width=\textwidth]{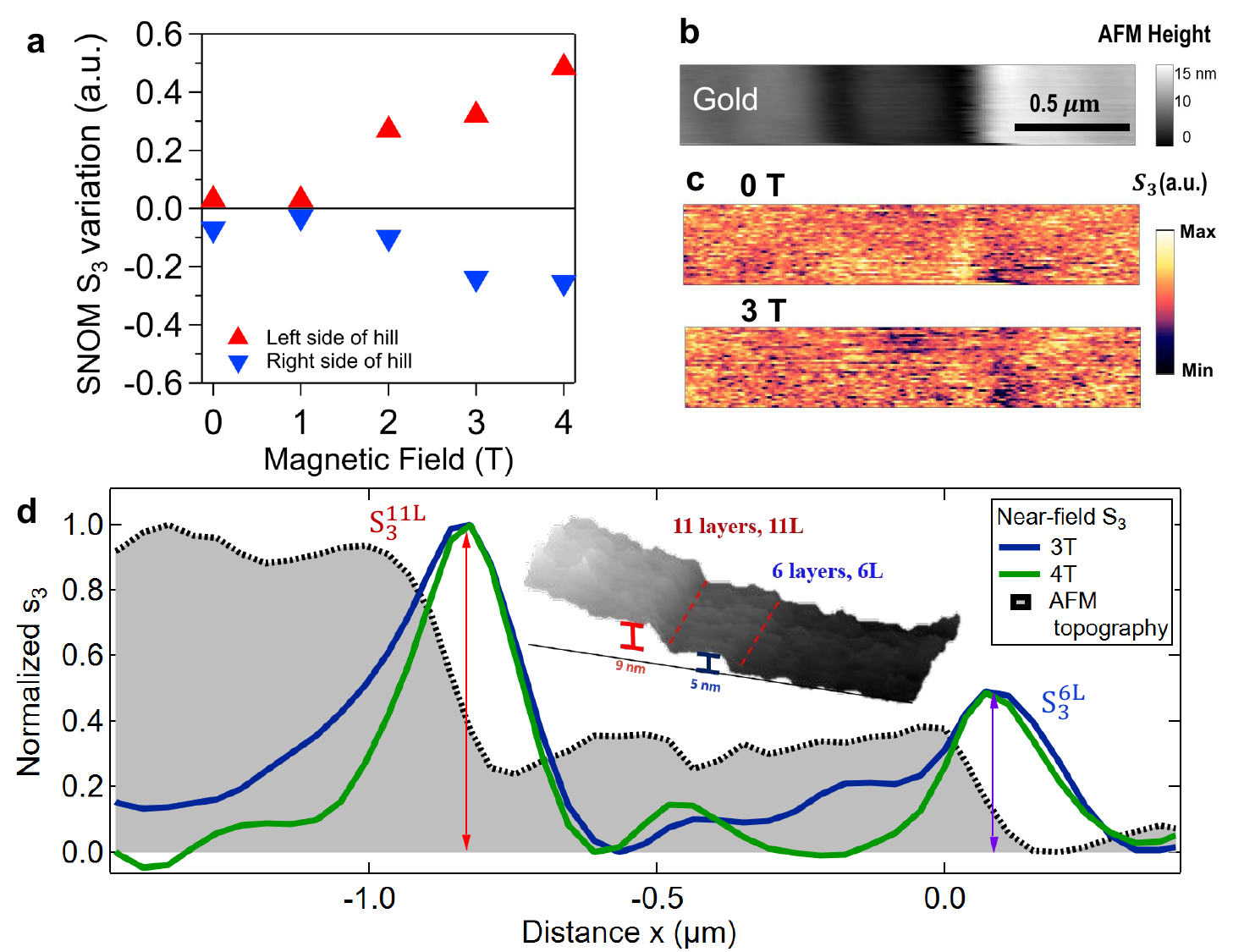} 
\caption{\textbf{Magnetic-field and layer-number dependence of near-field edge mode signals in ZrTe$_5$ and reference gold films.}
\textbf{a}, Near-field signals on the right (red) and left (blue) sides of the boundary, plotted as the maximum deviation from the local mean, are consistent with previously reported STM spin splitting measurements~\cite{LiPRL}.
\textbf{b}, AFM topography; \textbf{c}, $s_3$ sSNOM measurements acquired at 0~T and 3~T on a 160~nm gold film deposited over ZrTe$_5$ for comparison.
\textbf{d}, $s_3$ sSNOM line scans at 3~T and 4~T, shown alongside the AFM topography line scan (black dashed line) across two step edges. \textit{Inset}, 3D rendering of the AFM topography illustrating the measured region.Note that red dashed lines indicate step heights of 11 and 5 layers and highlight the height variations of key surface features.}
%Red and blue dashed lines indicate step heights of 11 and 5 layers, respectively, and highlight the height variations of key surface features.} 
\label{fig4:main}
\end{figure}

We present cm-IR-sSNOM scans of a 160~nm gold film as a control experiment to verify the consistency of our observations and the reliable magnetic nano-imaging operation at sub-2K temperatures. The gold film was deposited on a ZrTe$_5$ substrate cut from the same crystal used in this study. The gold topography, shown in Fig.~4b, closely follows the step structure of the underlying ZrTe$_5$. Magnetic-field scans at 0~T and 3~T, shown in Fig.~4c, reveal little to no meaningful change across steps with topographies similar to those examined in the bare ZrTe$_5$ data. We further validate AFM performance through a Sobel-gradient alignment procedure, enabling quantitative characterization via spatial Fourier transforms and the root-mean-square deviation from the mean AFM scan (Supplementary Information). These results confirm fully consistent AFM operation across all magnetic fields and temperatures with the Si-based cantilevers used here. Notably, the magnetic brightening of edge modes observed in ZrTe$_5$ is absent in the overlaid Au control film.
	
In the weak topological insulator phase studied above, ZrTe\textsubscript{5} can be viewed as a stack of nearly decoupled QSH layers, each contributing two spin-momentum locked, edge conduction channels. Under a uniform magnetic field that activates spin-polarized edge channels, the number of edge conduction channels can be directly inferred by comparing step edges of different layer thicknesses. Figure~\ref{fig4:main}d reveals a clear correlation between the number of edge-state channels and the near-field scattering signals.
Specifically, to determine the number of atomic layers, we measure step-edge heights from the inset of Fig.~\ref{fig4:main}d, which presents a 3D AFM topographic view of the structure in Fig.~2. The AFM profile shows two distinct boundaries between ZrTe$_5$ layers, indicated by red dashed lines (inset, Fig.~\ref{fig4:main}d). Multiple scans yield step heights of $\sim 9$~nm and $\sim 5$~nm, corresponding to $\approx 11$ and $\approx 6$ stacked layers, respectively. In Fig.~\ref{fig4:main}d, these two step edges are analyzed at high enough magnetic fields of 3~T (blue) and 4~T (green) to lift the degeneracy. The near-field signal amplitude scales proportionally with the number of edge conduction channels, with an amplitude ratio of $S_{3}^{6L}$/$S_{3}^{11L}$ is very close to 6/11 for the $\approx 6$-layer and $\approx 11$-layer steps---demonstrating good agreement with nearly linear atomic-layer scaling.
These findings are consistent with earlier STM observations and calculations \cite{wuPRX} that edge states in a single layer ZrTe\textsubscript{5} remain robust on top of thick slabs, despite weak van der Waals interlayer coupling, and that multiple edges can be effectively stacked along the out-of-plane direction. 
As further elaborated below using our simulations of ZrTe\textsubscript{5} in Fig.~5, higher step edges comprising two (or a few) atomic layers preserve the edge states, with a proportionally increased number of channels.
%These results strongly support the quantization of edge conduction channels in ZrTe\textsubscript{5} and are consistent with earlier STM observations ofZrTe\textsubscript{5} edge states.
%The ratio of near-field responses from simultaneously measured nanometer-scale step edges provides compelling evidence for the quantized scaling of edge conduction channels.

Figure~4d directly visualizes the spatial extent of magnetically separated helical edge modes which is governed by two key factors: the intrinsic localization of edge states and the magnetic confinement of evanescent electromagnetic modes. The intrinsic edge-state width is set by gapless local density of states which have been experimentally shown to be $\sim 6$~nm in ZrTe$_5$ from STM measurements near monolayer step boundaries \cite{wuPRX}. While this defines the quantum confinement of the underlying edge modes, the measured optical contrast cm-IR-sSNOM arises primarily from the electromagnetic response---specifically, the decay length of near-field evanescent modes localized at edges. This optical confinement is naturally linked to the Dirac mass gap $\Delta$ via the tunneling length scale $\lambda_\Delta = 2\hbar v_\text{F} / \Delta$, which sets the spatial decay of edge polarizability into surface gaps. For $\Delta \approx 6\text{–}10\,\text{meV}$ and $v_\text{F} \approx 5 \times 10^5\,\text{m/s}$, we estimate $\lambda_\Delta \approx 70\text{–}100\,\text{nm}$. This 
%gap-limited, field-independent scale, together with 5--10 layer step-edge 
broadening of our near-field electromagnetic response, together with tip radius convolution, qualitatively accounts for the observed edge-mode width of $\sim$100nm in our cm-IR-sSNOM images. Furthermore, the magnetic field further confines spin-split edge modes as illustrated in Figs.~1e-1g. Its effects manifest primarily in enhancing magnetic brightening due to spin population imbalance and slightly modifying the spatial extent of edge modes. 

%The first effect dominates the second, as will be shown below. First, the cyclotron radius, $r_\text{c} = E_\text{F} / (e B v_\text{F})$, with $E_\text{F} \approx 5\text{–}10\,\text{meV}$ in our samples used, defines an upper bound for the orbital extent and yields $r_\text{c} \sim 200\text{–}400\,\text{nm}$ at $B = 5\,\text{T}$. Although $r_\text{c}$ typically exceeds the intrinsic confinement scale $\lambda_\Delta$, its decrease with increasing field brings it into proximity with structural features such as domain wall widths and step-edge broadening ($\sim 100\,\text{nm}$). This spatial resonance between magnetic orbit size and structural confinement enhances edge-induced localization and contributes to the observed magnetic brightening of chiral edge modes. Second, although the magnetic confinement estimated from the semiclassical radius of curvature is expected to further localize edge states, this effect remains relatively modest because $r_\text{c} << \lambda_\Delta$. These conclusions are consistent with our near-field imaging results in Fig.~4d, where the 4~T edge profile is noticeably narrower than the 3~T one but shows only a modest $\sim 10$~\% on the confinement side, consistent with the behaviors illustrated in Figs.~1e-1g. 

\begin{figure}[!ht] 
\centering 
\includegraphics[width=\textwidth]{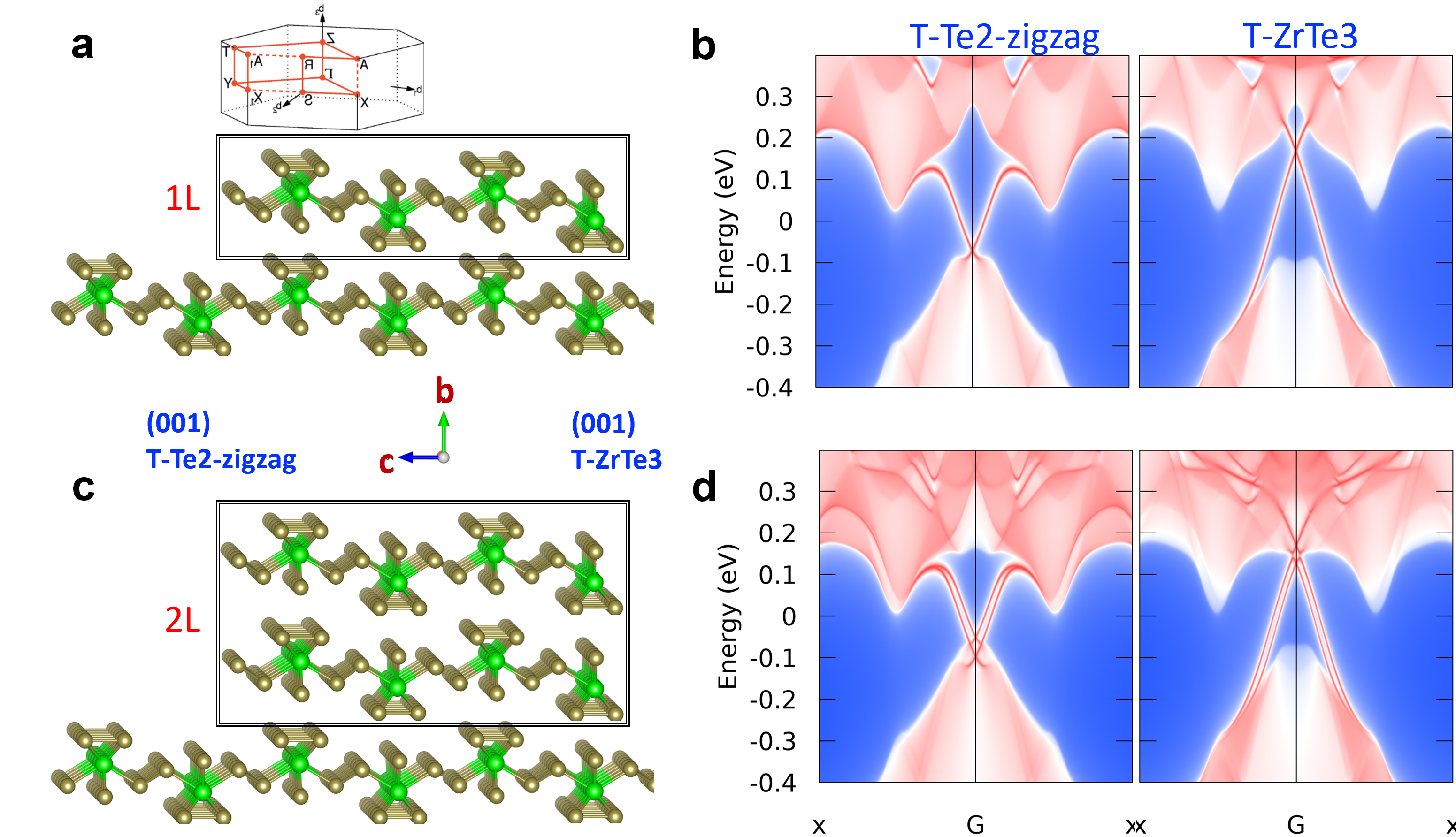} 
\caption{\textbf{Theoretical understanding of the band structure of ZrTe$_5$ in the context of a quantum spin Hall insulator for different edge terminations and layer numbers.} 
\textbf{a}, Illustration of a single step edge for 1 layer.
\textbf{b}, Band-structure calculations for monolayer ZrTe$_5$ with different edge terminations: telluride-terminated T-Te$_2$ zigzag (left) and zirconium-terminated T-ZrTe$_3$ (right).
\textbf{c}, Illustration of a single step edge for 2 layers. 
\textbf{d}, Band-structure calculations for bilayer ZrTe$_5$ with different edge terminations: telluride-terminated T-Te$_2$ zigzag (left) and zirconium-terminated T-ZrTe$_3$ (right).
} 
\label{fig5:main}
\end{figure}

To understand why magnetic-field-induced gaps do not significantly hybridize individual-layer edge states at the infrared energy scale, we perform band-structure calculations of ZrTe$_5$ in the context of QSH edge states. Conventional discussions of topological insulators typically focus on a single, isolated layer, which on its own exhibits the QSH effect. However, in realistic experiments of step edges in the samples studied, few layers can stack to produce atomically sharp edges spanning 10–20~nm in height, similar to the domain walls (Fig.~3) and step edges (Fig.~2). Such stacked layers may interact, modifying the single-layer band structure and giving rise to a more complex system. Figure~5a illustrates a single layer of ZrTe$_5$ with a truncated lattice, forming a sharp edge along the (001) direction where the crystal is most likely to cleave—namely, at the telluride atom. This is shown for monolayer ZrTe$_5$ for two distinct edge terminations: telluride-terminated T-Te$_2$ zigzag (left) and zirconium-terminated T-ZrTe$_3$ (right). In this simple single-layer case, our band-structure calculations in Fig.~5b reveal the emergence of the characteristic conical dispersion with a Dirac point at the edge, hosting a pair of counter-propagating helical edge states. Each of these contributes one quantum of conductance, $\frac{e^2}{h}$, consistent with the well-established QSH state and in agreement with earlier calculations~\cite{WengPRX} and STM studies \cite{LiPRL, wuPRX}.
	
We then extend the single-layer case to multiple layers terminating at the same lattice plane, giving rise to the experimentally observed atomically sharp, multi-layer edges. Figures~5c and 5d describe a two-layer edge of ZrTe$_5$, analogous to the single-layer configuration. Interestingly, at this two-layer edge, the conical band of the first layer remains largely intact, while a second conical band-slightly shifted in energy emerges, contributing its own pair of helical edge states. This behavior can be attributed to the weak van der Waals coupling between adjacent layers, which distorts the individual band structures and produces overlapping but shifted conical edge dispersions. Such weak interlayer coupling supports the notion of a robust edge state persisting across the ZrTe$_5$ terrace, as reported in previous STM works~\cite{LiPRL, wuPRX, WengPRX}.

A key feature of the ZrTe$_5$ system is the preservation of spin–momentum–locked helical edge states, whose number doubles in the two-layer case, thereby increasing the available spin channels within the step-edge region. We anticipate that this effect extends to higher-layer terminations, where stacking further increases the number of spin channels and, consequently, the total conductance at taller edges. This expectation is supported by our experimentally observed scaling behavior of near-field signals in Fig.~4d, which reveal that the conductance of multi-layer edges increases substantially---approximately scaling linearly with the number of atomic layers.

%The \textbf{quantum spin Hall (QSH)} effect can be viewed as the spin analog of the quantum Hall effect, featuring helical edge states protected by \textit{time-reversal symmetry}. In contrast, the \textbf{quantum anomalous Hall (QAH)} effect---also referred to as the \textit{Chern insulator phase}---is the charge analog, arising from spontaneous magnetization and characterized by \textit{broken time-reversal symmetry}. These two topological phases are distinguished by their respective topological invariants: $\mathbb{Z}_2$ for QSH and the \textit{Chern number} for QAH.

Our first-principles calculations in Fig.~\ref{fig5:main}, based on relaxed lattice parameters, show that ZrTe\textsubscript{5} in the \textit{weak topological insulator} regime does not host a surface Dirac cone on the terrace of the (010) surface imaged in the experimental configuration. This again agrees with earlier STM experiments \cite{LiPRL, wuPRX}. Consequently, a Chern-insulating edge state is not expected to emerge under an external magnetic field on this surface. Instead, the weak topological insulator feature of ZrTe\textsubscript{5} produces nanometer-scale step edges that can be viewed as stacks of QSH layers, each contributing a robust helical edge state. When a finite magnetic field is applied, the spin degeneracy of these edge states is lifted, producing a \textit{spin imbalance}. This imbalance generates a spin-polarized infrared polarizability and conductivity, which in turn manifests as the observed infrared near-field scattering signals.
%QAH
%In our infrared (IR) near-field imaging experiments, we observe a striking brightening of nanometer-scale step edges under applied magnetic fields--an effect that contrasts sharply with the quenching of edge conductivity reported in microwave (MW) and DC transport studies. The suppression in those lower-energy probes has been attributed to magnetic gap opening and inter-layer hybridization, arising from broken time-reversal symmetry and coupling between layers.

%More intriguingly, our IR near-field signals exhibit an approximately linear scaling with the number of layers, providing direct evidence that the magnetic field–induced gap is not large enough to disrupt or strongly hybridize the edge channels associated with each individual layer. On the energy scale of the $\sim$100 meV IR probe, the quantized edge conduction from each layer remains largely independent and robust, behaving as if the layers remain decoupled in terms of edge response.

%We therefore conclude that the magnetic gap is too small compared to the IR probe energy ($\sim$100 meV) to cause significant hybridization or destruction of the edge modes. The result is a near–layer-by-layer additive conductivity, with each layer effectively retaining its identity as a quantum spin Hall (QSH) channel at the measurement energy. This highlights the energy-scale-dependent sensitivity of edge-state transport and underscores the power of IR near-field techniques to probe topological edge responses in layered weak TIs such as ZrTe\textsubscript{5}.

\section{Summary}\label{sec4}

Nanoscale helical edge states remain robust under magnetic fields when probed at infrared energies in the weak topological insulator ZrTe\textsubscript{5}. Magneto–infrared near-field imaging reveals field-induced brightening and confinement of spin-polarized edge modes, with signal strength scaling with spin-split density of states and nearly linearly with layer number. These results show that the magnetic gap is far smaller than the $\sim 100$~meV probe energy, preserving independent quantum spin Hall identities in each layer. This tunable, energy-resilient edge transport highlights magnetic near-field techniques as a route to identifying and engineering high-frequency, low-loss conduction channels in topological quantum materials for next-generation microelectronics, spintronics and quantum information science.

\pagebreak

\bmhead{Acknowledgments} Work at Ames National Laboratory was supported by the U.S. Department of Energy (DOE), Basic Energy Sciences, Division of Materials Sciences \& Engineering, under Contract No. AC02-07CH11358. Some of the computation used resources of the National Energy Research Scientific Computing Center (NERSC), a DOE Office of Science User Facility.
The work at BNL was supported by the US Department of Energy, office of Basic Energy Sciences, contract no. DOE-SC0012704.

\bmhead{Author Contributions} 
S.H., R.H.J.K, R.K.C and J.M.K. performed the IR imaging measurements. 
G.G., P.L.M. and Q.L. developed the sample development and performed transport characterizations. 
L.L.W and T.K. developed the model with the help of J.W., M.M., and L.L.
L.-L.W. performed DFT calculations. 
J.W. and S.H. analyzed the SNOM data with the input of all authors. 
The paper is written by J.W. and S.H. with discussions from all authors. 
J.W. conceived and supervised the project.

	\bibliography{BibTex}

@article{hill,
	
	author = {R. Hillenbrand and R. Taubner and F. Keilmann},
	
	journal = {Nature},
	
	pages = {159--162},
	
	title = {Phonon-enhanced light–matter interaction at the nanometre scale},
	
	volume = {418},
	
	year = {2002},
	
	doi = {10.1038/nature00899},
	
}

@article{fei,
	
	author = {Z. Fei and A. S. Rodin and G. O. Andreev and W. Bao and A. S. McLeod and M. Wagner and L. M. Zhang and Z. Zhao and M. Thiemens and G. Dominguez and M. M. Fogler and A. H. Castro Neto and C. N. Lau and F. Keilmann and D. N. Basov},
	
	journal = {Nature},
	
	pages = {82--85},
	
	title = {Gate-tuning of graphene plasmons revealed by infrared nano-imaging},
	
	volume = {487},
	
	year = {2012},
	
	doi = {10.1038/nature11253},
	
}

@article{htchen,
	
	author = {H.-T. Chen and R. Kersting and G. C. Cho},
	
	journal = {Applied Physics Letters},
	
	pages = {3009--3011},
	
	title = {Terahertz imaging with nanometer resolution},
	
	volume = {83},
	
	year = {2003},
	
	doi = {10.1063/1.1616668},
	
}

@article{ribb,
	
	author = {H.-G. von Ribbeck and M. Brehm and D.W van der Weide and S. Winnerl and O. Drachenko and M. Helm and F. Keilmann},
	
	journal = {Optics Express},
	
	pages = {3430--3438},
	
	title = {Spectroscopic {TH}z near-field microscope},
	
	volume = {16},
	
	year = {2008},
	
	doi = {10.1364/oe.16.003430},
	
}

@article{kim,
	
	author = {Richard H. J. Kim and Chuankun Huang and Yilong Luan and Lin-Lin Wang and Zhaoyu Liu and Joong-Mok Park and Liang Luo and Pedro M. Lozano and Genda Gu and Deniz Turan and Nezih T. Yardimci and Mona Jarrahi and Ilias E. Perakis and Zhe Fei and Qiang Li and Jigang Wang},
	
	journal = {ACS Photonics},
	
	pages = {1873--1880},
	
	title = {Terahertz Nano-Imaging of Electronic Strip Heterogeneity in a {D}irac Semimetal},
	
	volume = {8},
	
	year = {2021},
	
	doi = {10.1021/acsphotonics.1c00216},
	
}

@article{kim2,
	
	author = {Richard H. J. Kim and Zhaoyu Liu and Chuankun Huang and Joong-Mok Park and Samuel J. Haeuser and Zhaoning Song and Yanfa Yan and Yongxin Yao and Liang Luo and Jigang Wang},
	
	journal = {ACS Photonics},
	
	pages = {3550--3556},
	
	title = {Terahertz Nanoimaging of Perovskite Solar Cell Materials},
	
	volume = {9},
	
	year = {2022},
	
	doi = {10.1021/acsphotonics.2c00861},
	
}

@article{kim3,
	
	author = {R. H. J. Kim and J.-M. Park and S. J. Haeuser and L. Luo and J. Wang},
	
	journal = {Review of Scientific Instruments},
	
	pages = {043702},
	
	title = {A sub-2 Kelvin cryogenic magneto-terahertz scattering-type scanning near-field optical microscope (cm-{TH}z-s{SNOM})},
	
	volume = {94},
	
	year = {2023},
	
	doi = {10.1063/5.0130680},
	
}

@article{kim4,
	
	author = {Richard H. J. Kim and Joong M. Park and Samuel Haeuser and Chuankun Huang and Di Cheng and Thomas Koschny and Jinsu Oh and Cameron Kopas and Hilal Cansizoglu and Kameshwar Yadavalli and Josh Mutus and Lin Zhou and Liang Luo and Matthew J. Kramer and Jigang Wang},
	
	journal = {Communications Physics},
	
	pages = {147},
	
	title = {Visualizing heterogeneous dipole fields by terahertz light coupling in individual nano-junctions},
	
	volume = {6},
	
	year = {2023},
	
	doi = {10.1038/s42005-023-01259-0},
	
}

@article{kim5,
	
	author = {Richard H. J. Kim and A. K. Pathak and Joong M. Park and Muhammad Imran and Samuel Haeuser and Zhe Fei and Yaroslav Mudryk and Thomas Koschny and Jigang Wang},
	
	journal = {Optics Express},
	
	pages = {},
	
	title = {Nano-compositional imaging of the lanthanum silicide system at THz wavelengths},
	
	volume = {},
	
	year = {2023},
	
	doi = {10.1364/OE.507414},
	
}

@article{mast,
	
	author = {S. Mastel and A. A. Govyadinov and T. V. A. G. de Oliveira and I. Amenabar and R. Hillenbrand},
	
	journal = {Applied Physics Letters},
	
	pages = {023113},
	
	title = {Nanoscale-resolved chemical identification of thin organic films using infrared near-field spectroscopy and standard {F}ourier transform infrared references},
	
	volume = {106},
	
	year = {2015},
	
	doi = {10.1063/1.4905507},
	
}

@article{vasw2,
	title={Light-driven Raman coherence as a nonthermal route to ultrafast topology switching in a Dirac semimetal},
	author={Vaswani, Chirag and Wang, L-L and Mudiyanselage, Dinusha Herath and Li, Q and Lozano, PM and Gu, GD and Cheng, Di and Song, Boqun and Luo, Liang and Kim, Richard HJ and others},
	journal={Physical Review X},
	volume={10},
	number={2},
	pages={021013},
	year={2020},
	publisher={APS}
}

@article{luo,
	
	author = {Liang Luo and Di Cheng and Boqun Song and Lin-Lin Wang and Chirag Vaswani and P. M. Lozano and G. Gu and Chuankun Huang and Richard H. J. Kim and Zhaoyu Liu and Joong-Mok Park and Yongxin Yao and Kaiming Ho and Ilias E. Perakis and Qiang Li and J. Wang},
	
	journal = {Nature Materials},
	
	pages = {329--334},
	
	title = {A light-induced phononic symmetry switch and giant dissipationless topological photocurrent in $\mathrm{ZrTe}_{5}$},
	
	volume = {20},
	
	year = {2021},
	
	doi = {10.1038/s41563-020-00882-4},
	
}

@article{von,
	title={Spectroscopic THz near-field microscope},
	author={Von Ribbeck, H-G and Brehm, M and Van der Weide, DW and Winnerl, S and Drachenko, O and Helm, M and Keilmann, F},
	journal={Optics Express},
	volume={16},
	number={5},
	pages={3430--3438},
	year={2008},
	publisher={Optica Publishing Group}
}

@article{dapolito,
	title={Scattering-type scanning near-field optical microscopy with Akiyama piezo-probes},
	author={Dapolito, Michael and Chen, Xinzhong and Li, Chaoran and Tsuneto, Makoto and Zhang, Shuai and Du, Xu and Liu, Mengkun and Gozar, Adrian},
	journal={Applied Physics Letters},
	volume={120},
	number={1},
	year={2022},
	publisher={AIP Publishing}
}

@article{haeuser,
	title={Analysis of Near-Field Magnetic Responses on ZrTe5 through Cryogenic Magneto-THz Nano-Imaging},
	author={Haeuser, Samuel and Kim, Richard HJ and Park, Joong-Mok and Chan, Randall K and Imran, Muhammad and Koschny, Thomas and Wang, Jigang},
	journal={Instruments},
	volume={8},
	number={1},
	pages={21},
	year={2024},
	publisher={MDPI}
}

@article{dapo,
	title={Infrared nano-imaging of Dirac magnetoexcitons in graphene},
	author={Dapolito, Michael and Tsuneto, Makoto and Zheng, Wenjun and Wehmeier, Lukas and Xu, Suheng and Chen, Xinzhong and Sun, Jiacheng and Du, Zengyi and Shao, Yinming and Jing, Ran and others},
	journal={Nature Nanotechnology},
	volume={18},
	number={12},
	pages={1409--1415},
	year={2023},
	publisher={Nature Publishing Group UK London}
}

@article{hillenbrand2025visible,
  title={Visible-to-THz near-field nanoscopy},
  author={Hillenbrand, Rainer and Abate, Yohannes and Liu, Mengkun and Chen, Xinzhong and Basov, Dmitri N},
  journal={Nature Reviews Materials},
  volume={10},
  number={4},
  pages={285--310},
  year={2025},
  publisher={Nature Publishing Group UK London}
}

@article{guo2024terahertz,
  title={Terahertz nanoscopy: Advances, challenges, and the road ahead},
  author={Guo, Xiao and Bertling, Karl and Donose, Bogdan C and Bruenig, Michael and Cernescu, Adrian and Govyadinov, Alexander A and Raki{\'c}, Aleksandar D},
  journal={Applied Physics Reviews},
  volume={11},
  number={2},
  year={2024},
  publisher={AIP Publishing}
}

@article{hasan,
	title = {Colloquium: Topological insulators},
	author = {Hasan, M. Z. and Kane, C. L.},
	journal = {Rev. Mod. Phys.},
	volume = {82},
	issue = {4},
	pages = {3045--3067},
	numpages = {0},
	year = {2010},
	month = {Nov},
	publisher = {American Physical Society},
	doi = {10.1103/RevModPhys.82.3045},
	url = {https://link.aps.org/doi/10.1103/RevModPhys.82.3045}
}

@article{Qi,
	title = {Topological insulators and superconductors},
	author = {Qi, Xiao-Liang and Zhang, Shou-Cheng},
	journal = {Rev. Mod. Phys.},
	volume = {83},
	issue = {4},
	pages = {1057--1110},
	numpages = {0},
	year = {2011},
	month = {Oct},
	publisher = {American Physical Society},
	doi = {10.1103/RevModPhys.83.1057},
	url = {https://link.aps.org/doi/10.1103/RevModPhys.83.1057}
}

@article{kons,
	title={Photoinduced dirac semimetal in ZrTe5},
	author={Konstantinova, T and Wu, L and Yin, W-G and Tao, J and Gu, GD and Wang, XJ and Yang, Jie and Zaliznyak, IA and Zhu, Y},
	journal={npj Quantum Materials},
	volume={5},
	number={1},
	pages={80},
	year={2020},
	publisher={Nature Publishing Group UK London}
}

@article{WengPRX,
	title = {Transition-Metal Pentatelluride $\mathrm{ZrTe}{}_{5}$ and $\mathrm{HfTe}{}_{5}$: A Paradigm for Large-Gap Quantum Spin Hall Insulators},
	author = {Weng, Hongming and Dai, Xi and Fang, Zhong},
	journal = {Phys. Rev. X},
	volume = {4},
	issue = {1},
	pages = {011002},
	numpages = {8},
	year = {2014},
	month = {Jan},
	publisher = {American Physical Society},
	doi = {10.1103/PhysRevX.4.011002},
	url = {https://link.aps.org/doi/10.1103/PhysRevX.4.011002}
}

@article{LiPRL,
	title = {Experimental Observation of Topological Edge States at the Surface Step Edge of the Topological Insulator ${\mathrm{ZrTe}}_{5}$},
	author = {Li, Xiang-Bing and Huang, Wen-Kai and Lv, Yang-Yang and Zhang, Kai-Wen and Yang, Chao-Long and Zhang, Bin-Bin and Chen, Y. B. and Yao, Shu-Hua and Zhou, Jian and Lu, Ming-Hui and Sheng, Li and Li, Shao-Chun and Jia, Jin-Feng and Xue, Qi-Kun and Chen, Yan-Feng and Xing, Ding-Yu},
	journal = {Phys. Rev. Lett.},
	volume = {116},
	issue = {17},
	pages = {176803},
	numpages = {5},
	year = {2016},
	month = {Apr},
	publisher = {American Physical Society},
	doi = {10.1103/PhysRevLett.116.176803},
	url = {https://link.aps.org/doi/10.1103/PhysRevLett.116.176803}
}

@article{wuPRX,
	title={Evidence for topological edge states in a large energy gap near the step edges on the surface of ZrTe 5},
	author={Wu, R and Ma, J-Z and Nie, S-M and Zhao, L-X and Huang, Xuejie and Yin, J-X and Fu, B-B and Richard, P and Chen, G-F and Fang, Zhong and others},
	journal={Physical Review X},
	volume={6},
	number={2},
	pages={021017},
	year={2016},
	publisher={APS}
}

@article{jutopo,
	title={Topological valley transport at bilayer graphene domain walls},
	author={Ju, Long and Shi, Zhiwen and Nair, Nityan and Lv, Yinchuan and Jin, Chenhao and Velasco Jr, Jairo and Ojeda-Aristizabal, Claudia and Bechtel, Hans A and Martin, Michael C and Zettl, Alex and others},
	journal={Nature},
	volume={520},
	number={7549},
	pages={650--655},
	year={2015},
	publisher={Nature Publishing Group UK London}
}

@article{shiMIM,
	title={Imaging quantum spin Hall edges in monolayer WTe2},
	author={Shi, Yanmeng and Kahn, Joshua and Niu, Ben and Fei, Zaiyao and Sun, Bosong and Cai, Xinghan and Francisco, Brian A and Wu, Di and Shen, Zhi-Xun and Xu, Xiaodong and others},
	journal={Science advances},
	volume={5},
	number={2},
	pages={eaat8799},
	year={2019},
	publisher={American Association for the Advancement of Science}
}

@article{Chen2015,
  author    = {Zhi-Guo Chen and Ruidan Y. Chen and Xiaofeng Lu and Yuan Zheng and Jianqiang Qi and Liang Wang and Fang Yang and Xianhui Chen and Zhaoguang Li and Jian-Hui Dai and H. K. Lee and N. L. Wang},
  title     = {Spectroscopic evidence for bulk-band inversion and three-dimensional massive Dirac fermions in ZrTe$_5$},
  journal   = {Proceedings of the National Academy of Sciences},
  volume    = {112},
  number    = {15},
  pages     = {4698--4702},
  year      = {2015},
  doi       = {10.1073/pnas.1421991112}
}

@article{Kim_new,
  author    = {Richard H. J. Kim and others},
  title     = {Terahertz Near-Field Imaging of Sidewall Losses in Superconducting Qubits},
  journal   = {Applied Physics Letters},
  year      = {2025},
  doi       = {10.1063/5.0284028}
}

@article{Jiang2023,
  author    = {Zhanzhi Jiang and Su Kong Chong and Peng Zhang and Peng Deng and Shizai Chu and Shahin Jahanbani and Kang L. Wang and Keji Lai},
  title     = {Implementing microwave impedance microscopy in a dilution refrigerator},
  journal   = {Review of Scientific Instruments},
  volume    = {94},
  number    = {5},
  pages     = {053701},
  year      = {2023},
  doi       = {10.1063/5.0148709}
}

@article{Eisele2014,
  author    = {M. Eisele and T. L. Cocker and M. A. Huber and M. Plankl and L. Viti and D. Ercolani and L. Sorba and M. S. Vitiello and R. Huber},
  title     = {Ultrafast multi-terahertz nano-spectroscopy with sub-cycle temporal resolution},
  journal   = {Nature Photonics},
  volume    = {8},
  pages     = {841--845},
  year      = {2014},
  doi       = {10.1038/nphoton.2014.163}
}
		
	%\pagebreak

\end{document}